# Temperature dependence and quenching characteristics of (La, Gd)$_2$Si$_2$O$_7$ scintillators at various Ce concentrations

*Masao Yoshino [a], Takahiko Horiai [a, d], Tatsuki Takasugi [a], Vitezslav Jary [d], Martin Nikl [d], Yuui Yokota [a], Kei Kamada [b, c], Yasuhiro Shoji [c], Romana Kucerkova [d], and Akira Yoshikawa [a, b, c]*

[a] Institute for Materials Research, Tohoku University, 2-1-1, Katahira, Aoba-ku, Sendai, Miyagi 982-8577, Japan

[b] New Industry Creation Hatchery Center, Tohoku University, 6-6-10, Aza-Aoba, Aramaki, Aoba-ku, Sendai, Miyagi 980-8579, Japan

[c] C&A Corporation, 6-6-40, Aza-Aoba, Aramaki, Aoba-ku, Sendai, Miyagi 980-8579, Japan

[d] Institute of Physics, Academy of Sciences of the Czech Republic, Cukrovarnicka 10, 162 00, Prague 6, Czech Republic

* Corresponding author. Email: yoshino.masao@tohoku.ac.jp

**Abstract**

We investigated the thermal stability of scintillation and the luminescence performances of (La, Gd)$_2$Si$_2$O$_7$ single crystals at various Ce concentrations. We prepared (La$_{0.25-x}$, Ce$_x$, Gd$_{0.75}$)$_2$Si$_2$O$_7$ (x = 0.0001, 0.001, 0.005, 0.01, 0.02, 0.05, and 0.1; unit: molar concentration) single crystals by the Czochralski and micro-pulling-down methods. With increasing Ce concentration, the photoluminescence emission and photoluminescence excitation spectral bands shifted to low energies and the activation energy $\Delta E$ for thermal quenching decreased. For Ce < 0.5 at. % samples, the photoluminescence emission background value calculated in the exponential approximation started to increase at temperatures greater than 320 K, which is probably because of Ce$^{3+}$ 5$d$ excited-state ionization. However, the effect was weaker for the Ce ≥ 0.5% samples, which may indicate a comparatively larger contribution from other nonradiative relaxations. Thus the main reason for the thermal quenching of the Ce$^{3+}$ emission in (La, Gd)$_2$Si$_2$O$_7$ is the combination of the 5$d$1 excited-state ionization and nonradiative relaxation via thermally excited crossover from the 5*d* excited state to the 4*f* ground state. The temperature dependence of the scintillation light yield was similar irrespective of the Ce concentration, with Ce 1.0% exhibiting the best performance within the temperature range 300 K to 450 K.

**1. Introduction**

Inorganic scintillators are essential in oil well logging and are used to detect gamma rays from the natural radioactivity of shale formation [1]. Because scintillation detectors in logging instrument packages operate at temperatures greater than 150°C, the temperature dependence of the scintillation properties is the key parameter that determines detector performance. At present, NaI:Tl is the most commonly used single-crystal scintillator used at commercial well logging detectors because of its high light yield [2]. However, NaI:Tl is hygroscopic and requires a hermetic seal, which reduces the useful scintillator volume available. The $Gd_2SiO_5$:Ce (GSO:Ce) scintillator, which is also used in oil well logging, is not hygroscopic and exhibits adequate light yield ca. 5000 ph/MeV at high temperatures [3]. However, the GSO:Ce single crystal has a (100) cleavage plane and a [001] slip direction in the cleavage plane [4], and thus packaging GSO:Ce requires additional space that one could otherwise use to enlarge the crystal.

In the past 2 dec, researchers have introduced Ce-doped pyrosilicate scintillators [$Lu_2Si_2O_7$:Ce (LPS:Ce) and $Gd_2Si_2O_7$:Ce (GPS:Ce)] [5, 6]. These pyrosilicate scintillators exhibit much higher light yield than GSO:Ce, a brief decay time (~40 ns), and are nonhygroscopic. Both LPS:Ce and GPS:Ce exhibit excellent thermal stability of light yield up to 450 K [7-9].

In 2012, Suzuki et al. [10] reported a cerium-doped (La, $Gd)_2Si_2O_7$ single-crystal scintillator, also known as LaGPS:Ce or La-GPS:Ce. (La, $Gd)_2Si_2O_7$:Ce exhibits an excellent light yield (36,000 ph/MeV), a fast scintillation decay time (46 ns), and high-energy resolution (5% at 662 keV). Kurosawa et al. [11] applied (La, Gd)$_2$Si$_2$O$_7$:Ce for oil well logging because of its high light yield (35,000 ph/MeV) and fast decay time (40 ns), even up to 150°C. Jary et al. [12] systematically compared the luminescence and scintillation characteristics of (La, Gd)$_2$Si$_2$O$_7$:Ce with the previously studied LPS:Ce. They also determined the barrier height of the thermal ionization of the Ce$^{3+}$ excited center and constructed an energy level diagram of the (La, Gd)$_2$Si$_2$O$_7$:Ce system. Murakami et al. [13] reported a crystal structure analysis of (La, Gd)$_2$Si$_2$O$_7$:Ce. Furthermore, Horiai et al. [14] investigated the divalent ion co-doping effect on (La, Gd)$_2$Si$_2$O$_7$:Ce. Most recently, Takasugi et al. [15] investigated the Al-doping effects on the mechanical, optical, and scintillation properties of (La, Gd)$_2$Si$_2$O$_7$:Ce [15]. Despite extensive work on the fundamental characteristics of (La, Gd)$_2$Si$_2$O$_7$:Ce scintillators, no systematic research has evaluated the Ce concentration effect on the scintillation and luminescence properties of (La, Gd)$_2$Si$_2$O$_7$. For Ce-activated scintillators, the Ce concentration is a crucial parameter that determines the scintillation efficiency and decay time. Additionally, the importance of Ce optimization increases when one uses the scintillators in a high-temperature environment. There is a downward shift of the quenching temperature for highly Ce-doped Y$_3$Al$_5$O$_{12}$ and Gd$_3$Ga$_3$Al$_2$O$_{12}$ [16, 17]. This shift is attributable to thermally activated concentration quenching [18]. The literature reports various proposed mechanisms that can be the origin of the thermal quenching of Ce$^{3+}$ $5d$–$4f$ emission. Two major mechanisms are as follows: (1) direct radiationless relaxation from the $5d$ excited state to the $4f$ ground state via a thermally activated crossover [19, 20], and (2) thermal ionization by excitation of the Ce$^{3+}$ electron from the $5d$ level to the conduction band [21, 22]. Jary et al. [12] indicated the presence of thermal ionization in Ce 1.5%-doped (La, Gd)$_2$Si$_2$O$_7$ [12].

We investigated the Ce concentration effects on the temperature quenching and concentration quenching in the (La, Gd)$_2$Si$_2$O$_7$:Ce system. We prepared single crystals of (La, Gd)$_2$Si$_2$O$_7$ with Ce percentages of 0.01% and 0.1% by the Czochralski (Cz) method; and 0.5%, 1.0%, 2.0%, 5.0%, and 10.0% by the micro-pulling-down (µ-PD) method. We measured the luminescence characteristics of (La, Gd)$_2$Si$_2$O$_7$:Ce to study the thermal quenching mechanisms. In particular, we measured the (1) photoluminescence (PL) emission and PL excitation spectra, (2) temperature dependence of the PL decay, and (3) temperature dependence of scintillation light yield within the 300 K to 450 K temperature range to evaluate the concentration quenching effect on the scintillation properties.

## 2. Materials and methods

### 2.1. Sample preparation

For sample preparation, powder materials of Gd$_2$O$_3$, La$_2$O$_3$, SiO$_2$, and CeO$_2$ with 4N purity (Iwatani Corporation, Tokyo, Japan) were used as the starting materials. These starting materials were mixed at a stoichiometry of (La$_{0.25-x}$, Ce$_x$, Gd$_{0.75}$)$_2$Si$_2$O$_7$ as follows; x = 0.0001, 0.001, 0.005, 0.01, 0.02, 0.05, and 0.1; unit: molar concentration. The Cz method was used to grow Ce 0.01% and 0.1% crystals, and the μ-PD method to grow Ce 0.5%, 1.0%, 2.0%, 5.0%, and 10% crystals. An undoped (La, Gd)$_2$Si$_2$O$_7$ single crystal was used as a seed. (La, Gd)$_2$Si$_2$O$_7$:Ce single crystals were grown from the melt (1750°C) using an Ir crucible. The crystal growth was carried out with a high-frequency induction furnace under pure argon and a 2 vol.% oxygen mixed-gas atmosphere. The growth method affects the scintillation performance of the crystals. According to Ref. [15], the influence of the growth method (Cz versus μ-PD) on the scintillation yield for (La, Gd)$_2$Si$_2$O$_7$:Ce was less than 15%. The diameter of the grown crystals by the Cz (Ce 0.01% and 0.1% crystals) was 55 mm. The diameter of the grown crystals by the μ-PD was 3 mm for Ce 0.5%, 1.0%, and 2.0% crystals, and 2 mm for Ce 5.0% and 10% crystals. The thickness of the measurement samples was 1.0 mm; and the surface size was 5 mm × 5 mm for Cz crystals and ϕ 3 mm or ϕ 2 mm for μ-PD crystals, respectively.

### 2.2. Sample characterization

Radioluminescence (RL) spectra and PL decay curves were measured with a Mo X-ray tube (RL spectra, 40 kV, 15 mA; Seifert, Baker Hughes, Houston, TX, USA) or a nanosecond nanoLED pulsed light source (model N-300; HORIBA Ltd., Kyoto, Japan) as the excitation source, and a custom-made Spectrofluorometer (5000M Horiba Jobin Yvon; model 199 Edinburgh Instruments, Bensheim, Germany) for the measurements. A single-grating monochromator and a photon counting detector (TBX-04; HORIBA, Ltd., Kyoto, Japan) were used as the detection unit. The measured RL spectra were corrected for the spectral dependence of the detection sensitivity. A convolution procedure was applied to the decay curves to determine the true decay time (SpectraSolve software package D7403; Ames Photonics, Wildwood, MA, USA). The PL characteristics in the temperature range of 77 K to 800 K were measured with a closed-cycle refrigerator (VPF Series Cryostat Systems VPF-800; Janis Instruments, Wildwood, MA, USA). For all RL spectra, $Bi_4Ge_3O_{12}$ (BGO) standard crystal scintillators were also measured under the same geometrical conditions to obtain quantitative intensity information. The PL decay time was calculated by fitting one or the sum of two exponentials:

$$A(t) = A_1 \exp(-t/\tau_1) + A_2 \exp(-t/\tau_2) + c, \qquad (1)$$

where $A_1, A_2$ and $\tau_1, \tau_2$ are the amplitudes and decay times, respectively, of the PL output.

Scintillation properties at room temperature were measured with a photomultiplier tube (PMT) (R6231-100; Hamamatsu Photonics K.K., Hamamatsu, Japan) under an 800-V bias voltage. A 1-MBq $^{137}$Cs source was placed 1 cm from the samples. $(La, Gd)_2Si_2O_7$:Ce samples were directly coupled to the PMT via OKEN 6262A silicone grease. The samples were covered with Teflon reflector tape. The scintillation decay time was calculated by fitting Eq. (1). For the scintillation decay measurements, the scintillator signal from the PMT was sent to a digital oscilloscope. The scintillation light yield was determined by comparing the photo-absorption peak of the $(La, Gd)_2Si_2O_7$:Ce samples and a reference sample [La-GPS(Ce), 5 mm × 5 mm × 5 mm, 45,000 photons/MeV; supplied by C&A Corp., Miyagi,

Japan]. For the light yield measurements, the signal from the PMT was fed into a spectroscopy amplifier (MSCF-16-F-V; Mesytec GmbH & Co., Putzbrunn, Germany), and then digitized using a list-mode Analogue to Digital Converter (ADC) (A3100; NIKI Glass Co. Ltd., Tokyo, Japan). The shaping time of the spectroscopy amplifier was set to 1 μs. For high-temperature measurements of the scintillation properties, we used an R1288AH PMT (Hamamatsu Photonics K.K., Hamamatsu, Japan) with a 1400-V bias voltage. High-temperature measurements were performed at 300 K, 375 K, 400 K, 425 K, and 450 K.

## 3. Results and discussion

### 3.1. PL emission, PL excitation, and temperature dependence of PL decay times

We first measured the PL emission and PL excitation spectra of (La, Gd)$_2$Si$_2$O$_7$:Ce samples at room temperature (Fig. 1). In the PL spectra of (La, Gd)$_2$Si$_2$O$_7$, all samples exhibited the typical 5$d$–4$f$ transition of Ce$^{3+}$ ca. 380 nm. We observed Gd$^{3+}$ 4$f$–4$f$ transitions at 310 and 275 nm in the PL excitation spectra of Ce ≤ 0.1% samples when monitoring the 380-nm emission of Ce$^{3+}$ 5$d$–4$f$ de-excitation. Bands in both the PL emission and PL excitation spectra shifted to lower energy with increasing Ce concentration.

Next, we measured the PL decays under 310 nm excitation [the 4$f$–5$d$ absorption band of Ce$^{3+}$ in (La, Gd)$_2$Si$_2$O$_7$] as a function of temperature (Fig. 2). For radiation measurements in high-temperature environments, the influence of the quenching temperature in (La, Gd)$_2$Si$_2$O$_7$:Ce is meaningful. Thermal quenching accelerates the PL decay time by an additional nonradiative pathway to the decay process and concomitantly lowers the scintillation light yield [23]. The decay time at 77 K was nonmonotonic and independent of the Ce concentration. At 77 K, the PL decay time was ca. 24 ns for the Ce 0.01% sample and ca. 27–30 ns for the Ce ≥ 0.1% samples under consideration. The increasing value of the decay time with temperature for heavily doped samples up to 400 K is probably because of reabsorption: researchers have described such an effect in LPS, Y$_3$Al$_5$O$_{12}$, YAlO$_3$, CaF$_2$, and YLiF$_4$ [7, 24, 25]; and it is even more pronounced for (La, Gd)$_2$Si$_2$O$_7$:Ce because of the comparatively smaller Stokes shift [12]. The rapid decrease of the PL decay times above 400 K, which was more evident for higher Ce concentrations, was a combined effect of thermal and concentration quenching—the latter acting at high Ce concentrations. The total decay rate is given by:

$$\tau_{\text{exp}}^{-1} = \tau_{\text{R}}^{-1} + \tau_{\text{NR}}^{-1}, \qquad (2)$$

where $\tau_{\text{exp}}$ is the experimental PL lifetime of the 5$d$–4$f$ transition; and $\tau_{\text{R}}$ and $\tau_{\text{NR}}$ are the contributions from the radiative and nonradiative processes, respectively. We calculated the nonradiative decay rate $W_{\text{NR}}$ ($= \tau_{\text{NR}}^{-1}$) from Eq. (2). We used the experimental PL decay rate at 77 K as the radiative decay rate $\tau_{\text{R}}^{-1}$. Fig. 3 shows the temperature dependence of the deduced nonradiative relaxation rate. We fit the plots assuming that $W_{\text{NR}}$ varies with temperature, and it follows an Arrhenius law:

$$W_{\text{NR}} = W_0 \exp\left(-\frac{\Delta E}{k_{\text{B}}T}\right), \qquad (3)$$

where $k_{\text{B}}$ is the Boltzmann constant, $W_0$ is the frequency factor, and $\Delta E$ is the activation energy. The roll-off temperature $T^*$ is the temperature at which the nonradiative decay rate becomes comparable to the radiative decay rate. Because the effect of self-absorption is greater when one uses data below the roll-off temperature, we restricted the fitting range to above the roll-off temperature. Table 1 and Fig. 1 summarize the fit parameters for the nonradiative decay process in the (La, Gd)$_2$Si$_2$O$_7$:Ce samples. We also plotted previously studied Ce 1.0%-doped (La$_{0.3}$, Gd$_{0.7}$)$_2$Si$_2$O$_7$ and (La$_{0.48}$, Gd$_{0.52}$)$_2$Si$_2$O$_7$ single cystals [12] (Fig. 4). Both the frequency factor $W_0$ and activation energy $\Delta E$ considerably decreased with increasing Ce concentration. We calculated the frequency factors to be $1.3 \times 10^{13}$ and $6.5 \times 10^9$ s$^{-1}$ for the Ce 0.01% and Ce 10% samples, respectively; the corresponding activation energies were 0.61 and 0.22 eV. As a result, the roll-off temperature $T^*$ slightly decreased with Ce concentration.

Fig. 5 shows the temperature dependence of the background intensity of the PL decay curves up to 500 K. All of the measurement conditions were unchanged. At greater than ~320 K, the background value calculated in the exponential fitting started to increase. This increase is most probably because of the $Ce^{3+}$ 5$d$ excited-state ionization, but the effect was weaker for the Ce ≥ 0.5% samples, which may indicate a larger contribution from other nonradiative relaxations. Alternatively, this might be because of the fast capture of an ionized electron in the conduction band by another $Ce^{4+}$ nearby; i.e., reduced trapping of these electrons by other traps. A further understanding of the ionization behavior would require an independent study of the characteristics of the traps involved in the delayed recombination [26].

Regarding the quenching mechanism of (La, Gd)$_2$Si$_2$O$_7$, the related results are as follows:

(a) Both PL emission and PL excitation spectra bands shifted to lower energy with increasing Ce concentration.

(b) The activation energy $\Delta E$ for the thermal quenching decreased with increasing Ce concentration.

(c) The roll-off temperature $T^*$ slightly decreased with increasing Ce concentration.

Jary et al. [12] measured the temperature dependence of the delayed recombination decay of Ce 1.0%-doped (La$_{0.3}$, Gd$_{0.7}$)$_2$Si$_2$O$_7$, and indicated that the nanosecond-scale reduction in the decay time is because of thermal ionization of the excited states. Our results regarding the temperature dependence of the background intensity of the PL decay (Fig. 5) are consistent with Jary et al.[12] The activation energy started to decrease at greater than 1.0% Ce. This phenomenon may indicate that at greater than 1.0% Ce, the contribution of the nonradiative relaxation that is attributable to thermal excitation crossover became larger with increasing Ce concentration.

Accordingly, the main reason for the thermal quenching of the $Ce^{3+}$ emission in (La, Gd)$_2$Si$_2$O$_7$ is the combination of the $5d$1 excited-state ionization and nonradiative relaxation via thermally excited crossover from the $5d$ excited state to the $4f$ ground state (Fig. 6).

### 3.2. Temperature dependence of scintillation properties

Fig. 7 shows the room-temperature spectra (X-ray 40 kV, 15 mA) of (La, Gd)$_2$Si$_2$O$_7$:Ce samples compared with that of a BGO reference. For the lowest Ce 0.01% concentration sample, there was an apparent high-energy part of the spectrum peaking ca. 340 nm, and the emission of $Gd^{3+}$ at 311 nm was also clearly visible. With increasing Ce concentration, there was increasing reabsorption at the high-energy side of the spectrum because of the comparatively smaller Stokes shift of $Ce^{3+}$ in this host [12]; accordingly, the registered spectrum low energy shifted by 30 nm measured at the half-height. Some effects that are attributable to the high concentration (such as $Ce^{3+}$ perturbed by another $Ce^{3+}$ nearby, change in the lattice constant, and $Ce^{3+}$ dimers) can affect the shift of the spectrum and cannot be excluded. We evaluated the scintillation efficiency by calculating the overall RL intensity. The overall RL intensity of Ce 0.01%, 0.1%, 1.0%, and 10% reached 1060%, 2400%, 2620%, and 2760%, respectively, compared with that exhibited by BGO. Table 2 summarizes the calculated RL intensity. At greater than Ce 0.1%, the RL intensity was nearly the same (2300% to 2600%).

Fig. 8 shows the energy spectra and scintillation decays of (La, Gd)$_2$Si$_2$O$_7$ with various Ce concentrations excited by a $^{137}$Cs source. The highest light yield was 40,000 ph/MeV, measured on a sample of 1.0% Ce. The scintillation decay time decreased with increasing Ce concentration. The values for the first component of the decay time were 95 and 52 ns for the Ce 0.1% and Ce 10% samples, respectively. Table 1 summarizes the results of calculations on the scintillation properties. Fig. 9 shows the relative RL intensity (× BGO) and scintillation light yield of (La, Gd)$_2$Si$_2$O$_7$ as a function of Ce concentration. For the Ce ≥ 0.5% samples, both the RL intensity and the light yield were within ±10% deviation and thus the concentration quenching did not considerably affect the scintillation efficiency for the (La, Gd)$_2$Si$_2$O$_7$ host; this result is consistent with that of the PL decay temperature dependence. The Ce 0.1% sample exhibited different trends in terms of the RL intensity and light yield. This difference may be because of the contribution of the relatively long microsecond- to millisecond-order 4$f$–4$f$ forbidden transition of Gd$^{3+}$. One can collect the slow luminescence that is attributable to the 4$f$–4$f$ transition of Gd$^{3+}$ by integrated RL intensity measurements, but not by scintillation light yield measurements. Fig. 10 shows the values and ratios of the primary and secondary decay component in the scintillation decays as a function of the Ce concentration. The slower secondary decay is attributable to the slow energy transfer from the ionizing radiation to the Ce$^{3+}$; in general, a possible explanation for Gd-containing scintillators is the energy migration among the Gd$^{3+}$ ions. Although we did not measure the Ce concentration, we assumed that the Ce$^{3+}$ concentration was proportional to that of the starting material. Thus both the values and ratios decreased with increasing Ce concentration. This result indicates that the average energy transfer rate from Gd$^{3+}$ to Ce$^{3+}$ increased with increasing Ce concentration.

Fig. 11 shows the temperature dependence of gamma-ray light yield at various Ce concentrations. We observed a similar trend for all of the samples; in agreement with the PL decay results, in which the quenching temperature did not appreciably depend on the Ce concentration. For applications in high-temperature environments up to 450 K, a Ce concentration ca. 1.0% was suitable for (La, Gd)$_2$Si$_2$O$_7$.

## 4. Conclusion

We investigated the temperature dependence of PL decays and scintillation light yield to understand the quenching effect for (La, Gd)$_2$Si$_2$O$_7$ in accordance with Ce concentration. We grew (La, Gd)$_2$Si$_2$O$_7$:Ce single crystals by the Cz and μ-PD methods. We measured the room-temperature PL emission and PL excitation spectra in the Ce 0.01%, 0.1%, 0.5%, 1.0%, 2.0%, 5.0%, and 10% samples. Among all of the (La, Gd)$_2$Si$_2$O$_7$ samples, both the PL emission and PL excitation spectra shifted to lower energy with increasing Ce concentration. We measured the PL decays between 77 K and 700 K. We observed an increasing decay time with temperature for heavily Ce-doped samples up to 400 K, most likely because of reabsorption. The onset of the shortening of the decay time appeared within the 400 K to 440 K temperature range and was less dependent on the Ce concentration. The activation energy $\Delta E$ and frequency factor $W_0$ decreased with increasing Ce concentration. Regarding the temperature dependence of the background in PL decays, at greater than 320 K the background value calculated in the exponential approximation started to increase, most likely because of excited-state ionization of Ce$^{3+}$. However, the background increase was weaker for higher-Ce samples, which may indicate a larger contribution from other nonradiative relaxations. Alternatively, this might be because of the fast capture of an ionized electron in the conduction band by another Ce$^{4+}$ nearby. Accordingly, the main reason for the thermal quenching of Ce$^{3+}$ emission in (La, Gd)$_2$Si$_2$O$_7$ is the combination of the 5$d$1 excited-state ionization and nonradiative relaxation via thermally excited crossover from the 5$d$ excited state to the 4$f$ ground state. Future work should include independent studies of the characteristics of the traps involved in the delayed recombination, to better understand the ionization. Measured room-temperature RL spectra (X-ray 40 kV, 15 mA) indicated that with increasing Ce concentration there was increasing reabsorption at the high-energy side of the spectrum because of the comparatively smaller Stokes shift of Ce$^{3+}$. The calculated RL intensity of the Ce 0.01%, 0.1%, 1.0%, and 10% samples reached 1060%, 2400%, 2620%, and 2760%, respectively, compared with that exhibited by BGO. Regarding the scintillation light yield, the trend of the temperature dependence was similar for all of the samples, and a Ce concentration of 1.0% exhibited the best performance within the 300 K to 450 K temperature range.

**Acknowledgments**

This work was supported by JSPS KAKENHI [grant numbers 21H03834, 19K17191]. Partial support of the Operational Program Research, Development, and Education financed by European Structural and Investment Funds and the Czech Ministry of Education, Youth, and Sports (Project No. SOLID21 CZ.02.1.01/0.0/0.0/16_019/0000760) is gratefully acknowledged; as is support by the bilateral Japanese–Czech mobility project. The sponsors had no role in study design; in the collection, analysis, and interpretation of data; in the writing of the report; and in the decision to submit the article for publication. We thank Michael Scott Long, PhD, from Edanz Group (https://jp.edanz.com/ac) for editing a draft of this manuscript.

Received: ((will be filled in by the editorial staff))
Revised: ((will be filled in by the editorial staff))
Published online: ((will be filled in by the editorial staff))

**Table 1. Summary of radiative and nonradiative decay parameters for (La, Gd)$_2$Si$_2$O$_7$:Ce samples.**

| Ce concentration (%) | $\tau_R$ (ns) | $W_0$ (s$^{-1}$) | $\Delta E$ (eV) | $T^*$ (K) |
|---|---|---|---|---|
| 0.01 | 24.9 | $1.3 \times 10^{13}$ | 0.61 | 555 |
| 0.1 | 28.7 | $1.5 \times 10^{13}$ | 0.64 | 572 |
| 0.5 | 30.7 | $1.3 \times 10^{13}$ | 0.63 | 568 |
| 1.0 | 29.7 | $1.9 \times 10^{12}$ | 0.50 | 536 |
| 2.0 | 28.6 | $3.2 \times 10^{11}$ | 0.39 | 499 |
| 5.0 | 27.8 | $4.6 \times 10^{11}$ | 0.43 | 526 |
| 10 | 27.5 | $6.5 \times 10^9$ | 0.22 | 484 |

**Table 2. Summary of scintillation performance for (La, Gd)$_2$Si$_2$O$_7$:Ce samples.**

| Ce concentration (%) | RL intensity (%, × BGO[a]) | LY (ph/MeV) | Decay (ns) 1st | Decay (ns) 2nd |
|---|---|---|---|---|
| 0.01 | 1,060 | not detected | 95 (20%) | 1200 (80%) |
| 0.1 | 2,400 | 20,000 | 71 (14%) | 620 (86%) |
| 0.5 | 2,620 | 37,000 | 61 (18%) | 370 (82%) |
| 1.0 | 2,620 | 40,000 | 68 (29%) | 230 (71%) |
| 2.0 | 2,330 | 38,000 | 69 (88%) | 240 (12%) |
| 5.0 | 2,410 | 35,000 | 52 (86%) | 390 (14%) |
| 10 | 2,760 | 33,000 | 45 (81%) | 300 (19%) |

[a]BGO: Bi$_4$Ge$_3$O$_{12}$; LY: light yield; RL: radioluminescence.

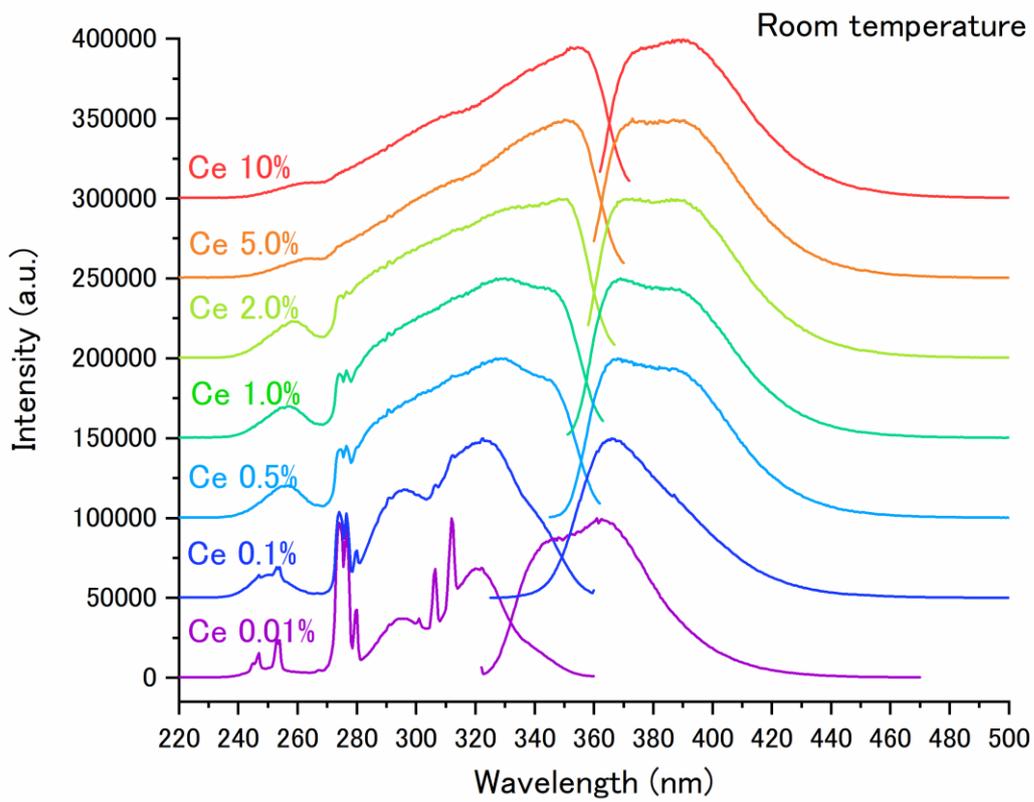

**Figure 1.** Photoluminescence emission (for excitation to the $5d_1$ state) and photoluminescence excitation (for $5d_1$–$4f$ luminescence) spectra of $(La, Gd)_2Si_2O_7$:Ce at various Ce concentrations.

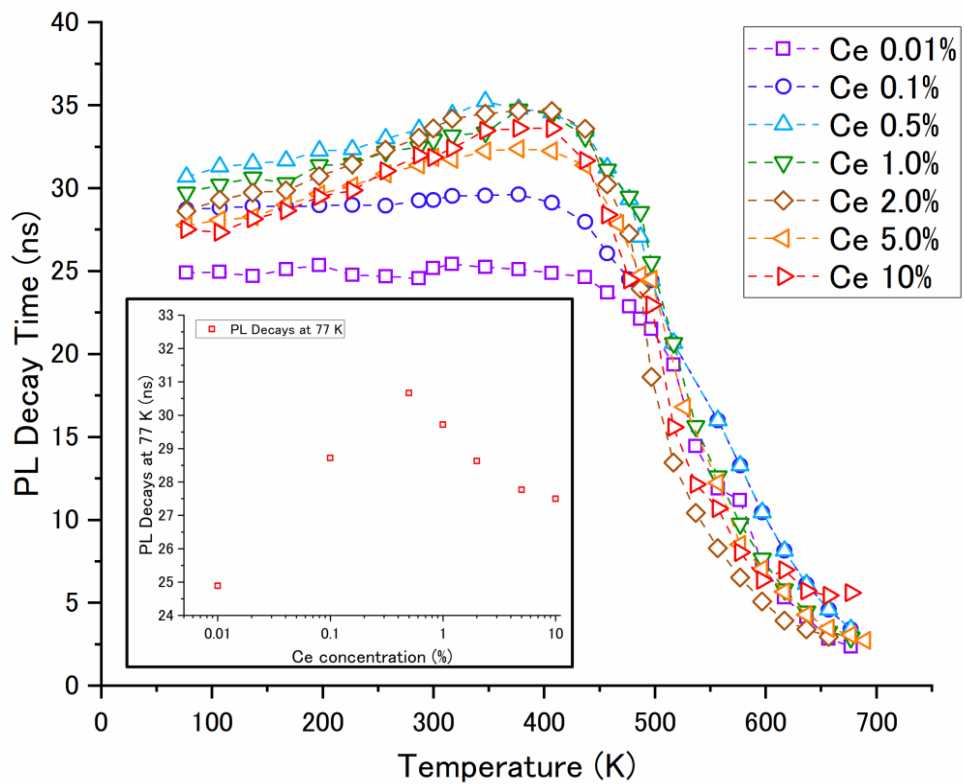

**Figure 2. Photoluminescence decay times determined from single-exponential fits as a function of temperature for (La, Gd)$_2$Si$_2$O$_7$:Ce at various Ce concentrations.**

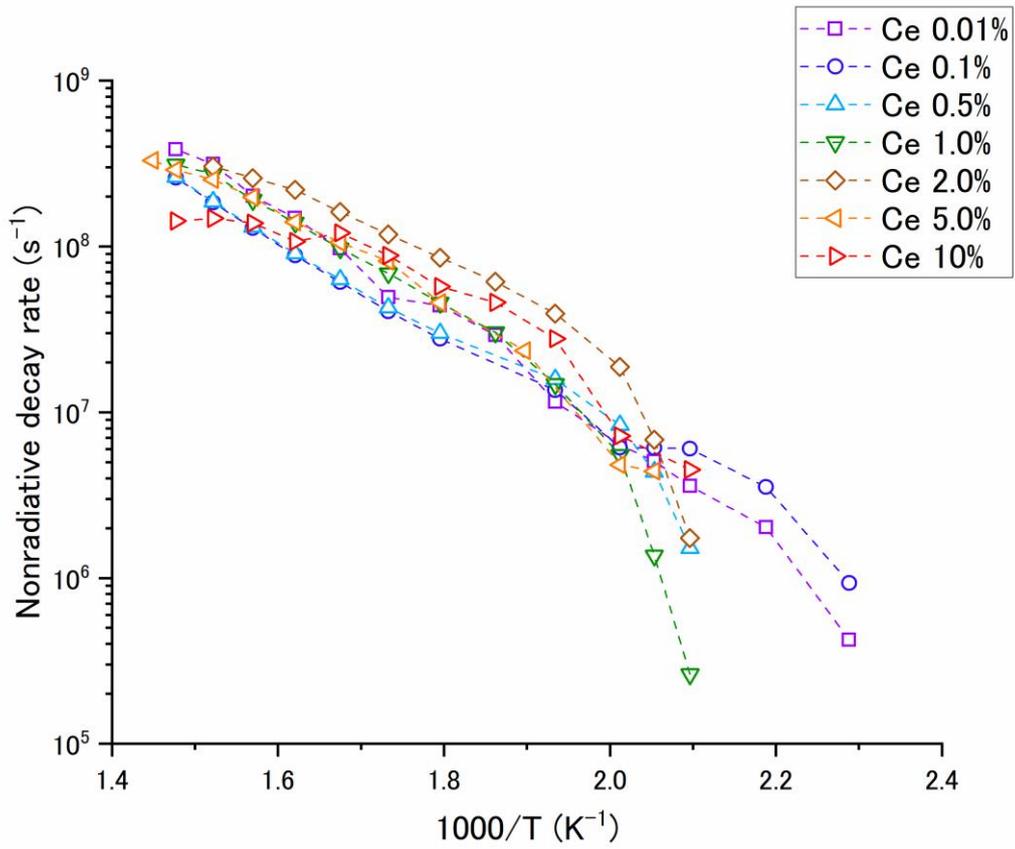

**Figure 3. Nonradiative decay rate for (La, Gd)$_2$Si$_2$O$_7$:Ce samples as a function of inverse temperature at various Ce concentrations.**

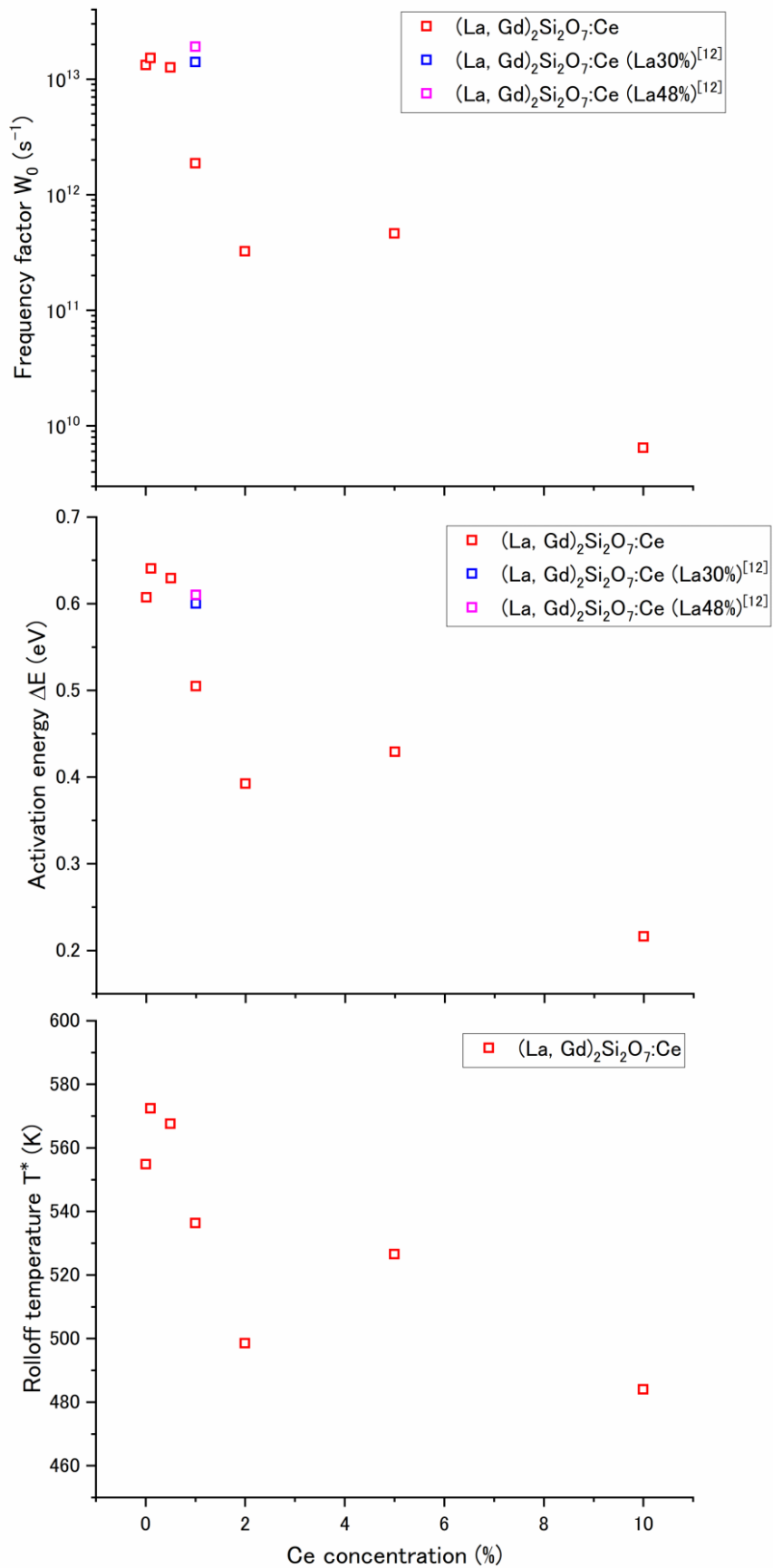

**Figure 4. Frequency factor, activation energy, and roll-off temperature of (La, Gd)$_2$Si$_2$O$_7$ as a function of Ce concentration.**

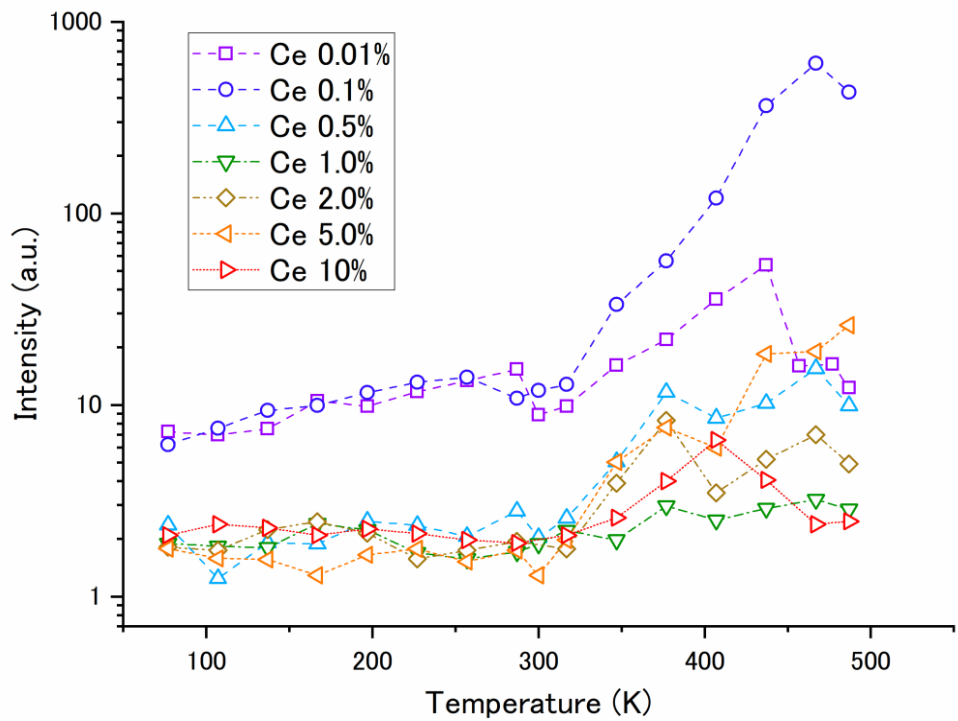

**Figure 5: Temperature dependence of the background intensity for (La, Gd)$_2$Si$_2$O$_7$:Ce at various Ce concentrations.**

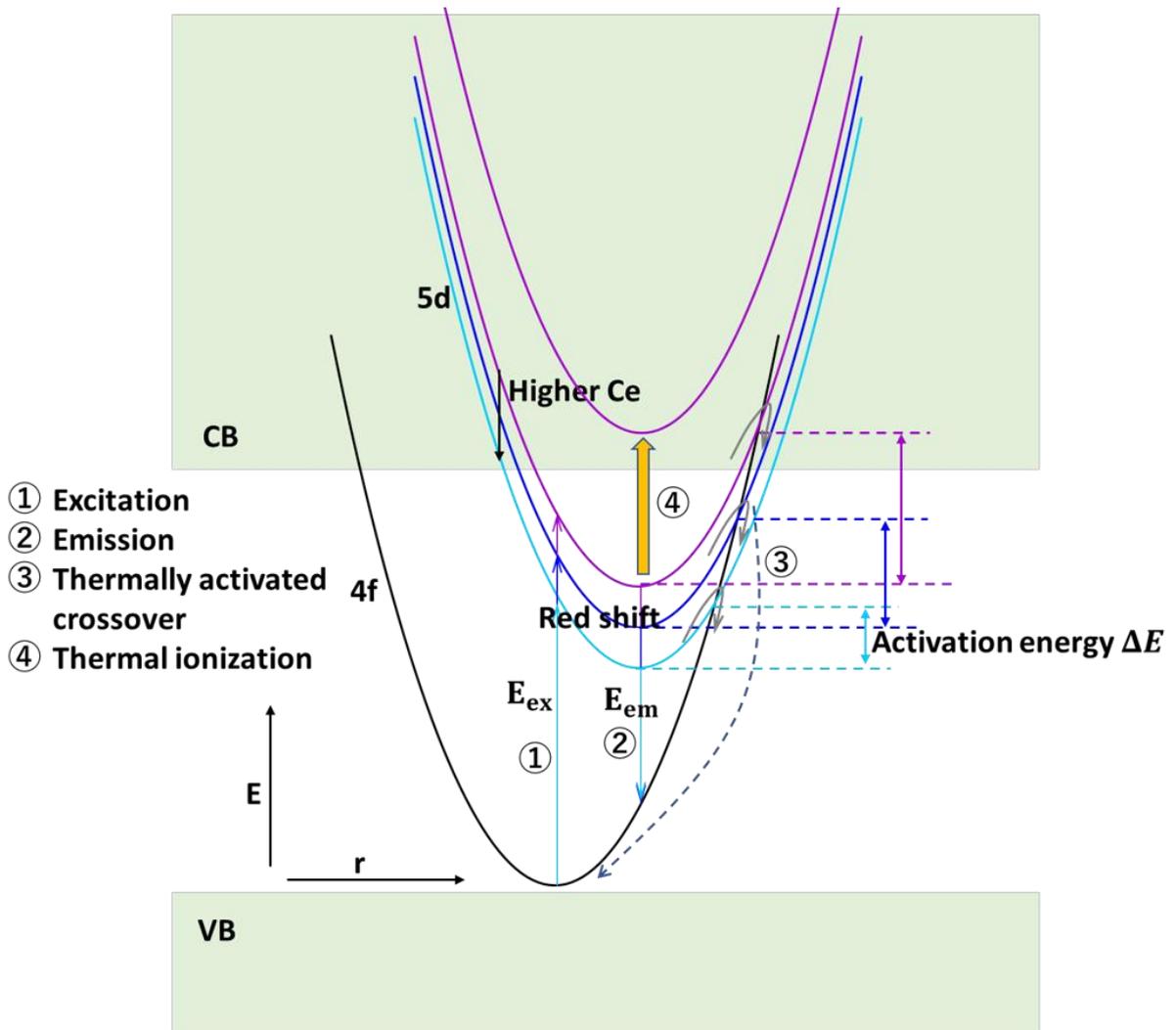

**Figure 6.** Configurational coordinate diagram of the 4*f* and 5*d* state of $Ce^{3+}$, with energy E versus configurational coordinate r in $(La, Gd)_2Si_2O_7$.

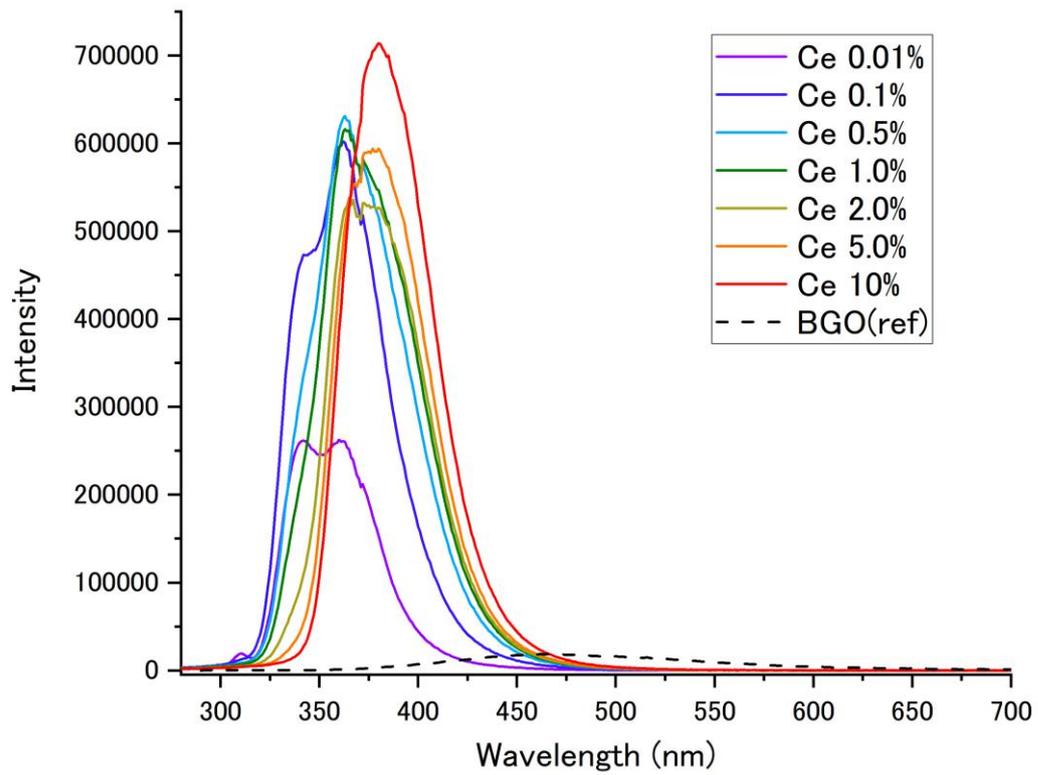

**Figure 7. Radioluminescence spectra of (La, Gd)$_2$Si$_2$O$_7$:Ce at various Ce concentrations, irradiated with an X-ray tube (40 kV, 15 mA). (BGO: Bi$_4$Ge$_3$O$_{12}$)**

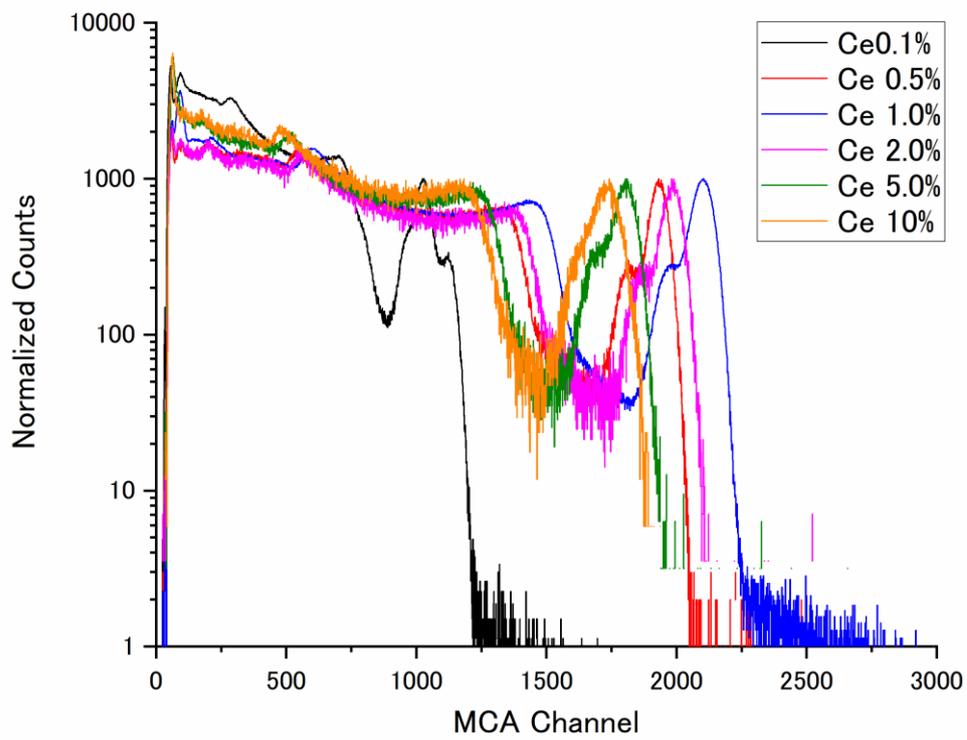

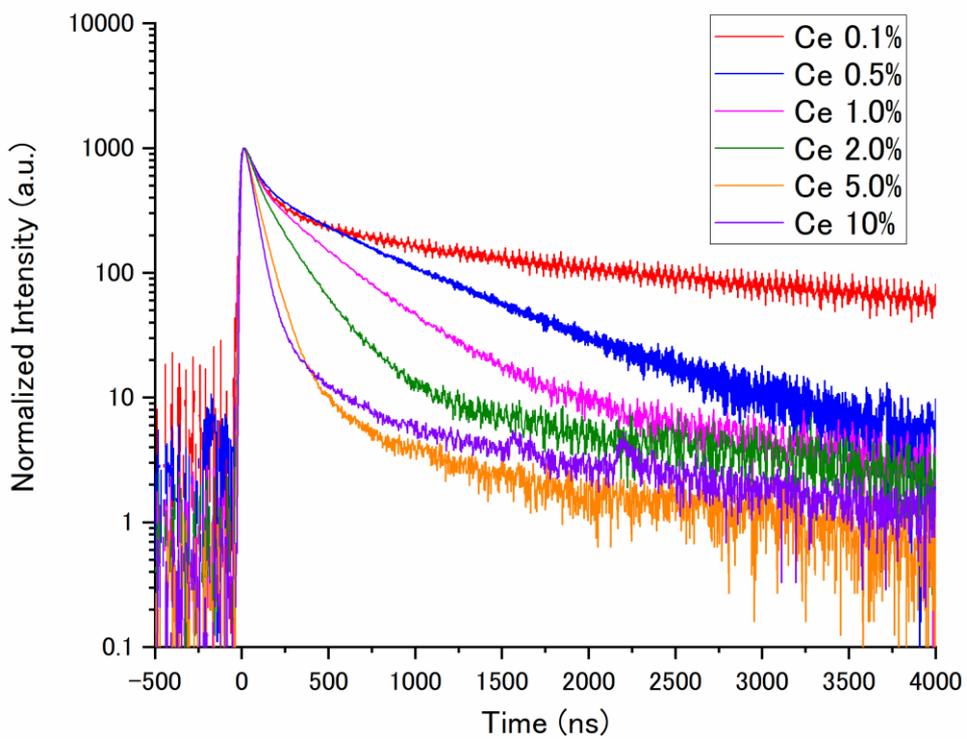

**Figure 8. Energy spectra and scintillation decay of (La, Gd)$_2$Si$_2$O$_7$ at various Ce concentrations, excited with a $^{137}$Cs source.**
MCA: multi channel analyzer

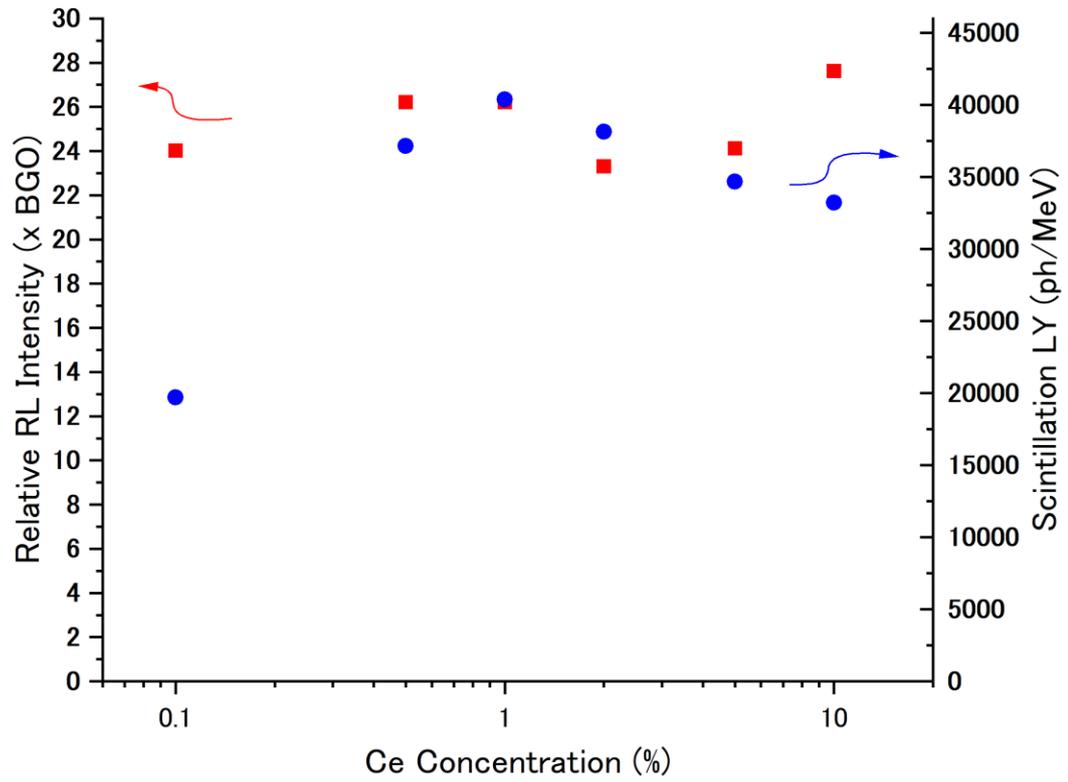

**Figure 9. Ce concentration vs. relative radioluminescence (RL) intensity [×Bi$_4$Ge$_3$O$_{12}$ (BGO)] and scintillation light yield (LY) of (La, Gd)$_2$Si$_2$O$_7$:Ce. Squares: left-hand axis. Circles: right-hand axis.**

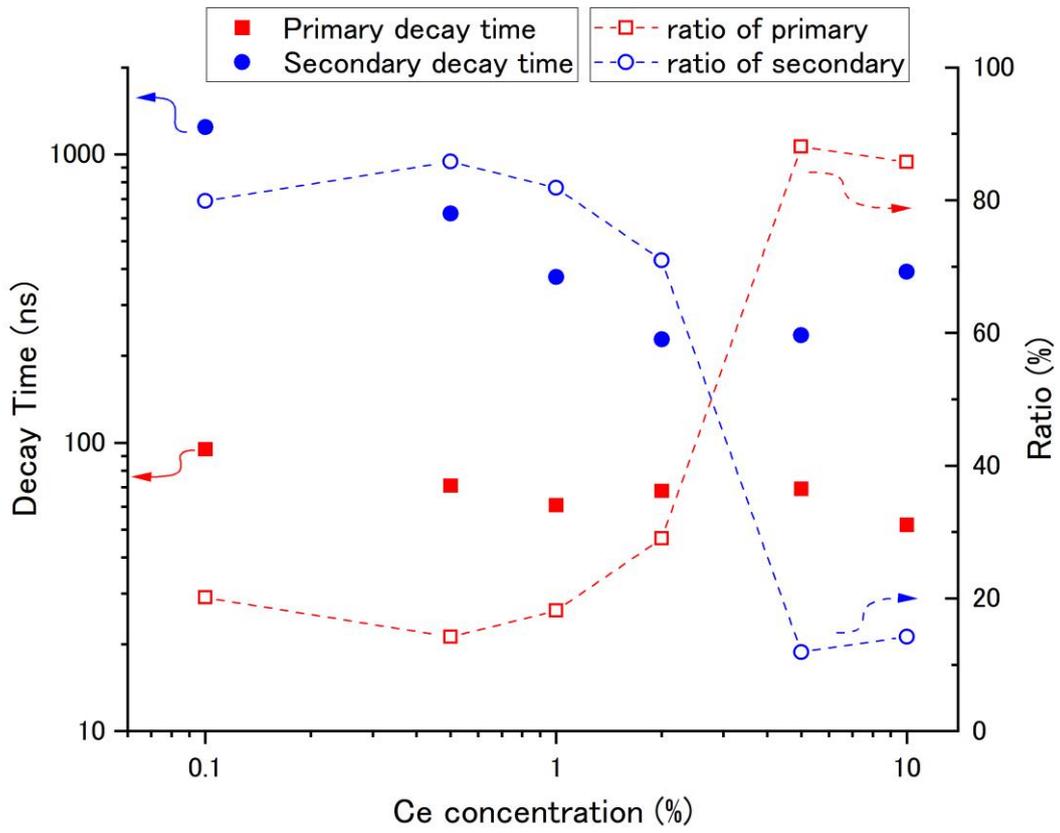

**Figure 10. Ce concentration vs. primary/secondary decay time and the ratio of primary/secondary decay in (La, Gd)$_2$Si$_2$O$_7$:Ce.**

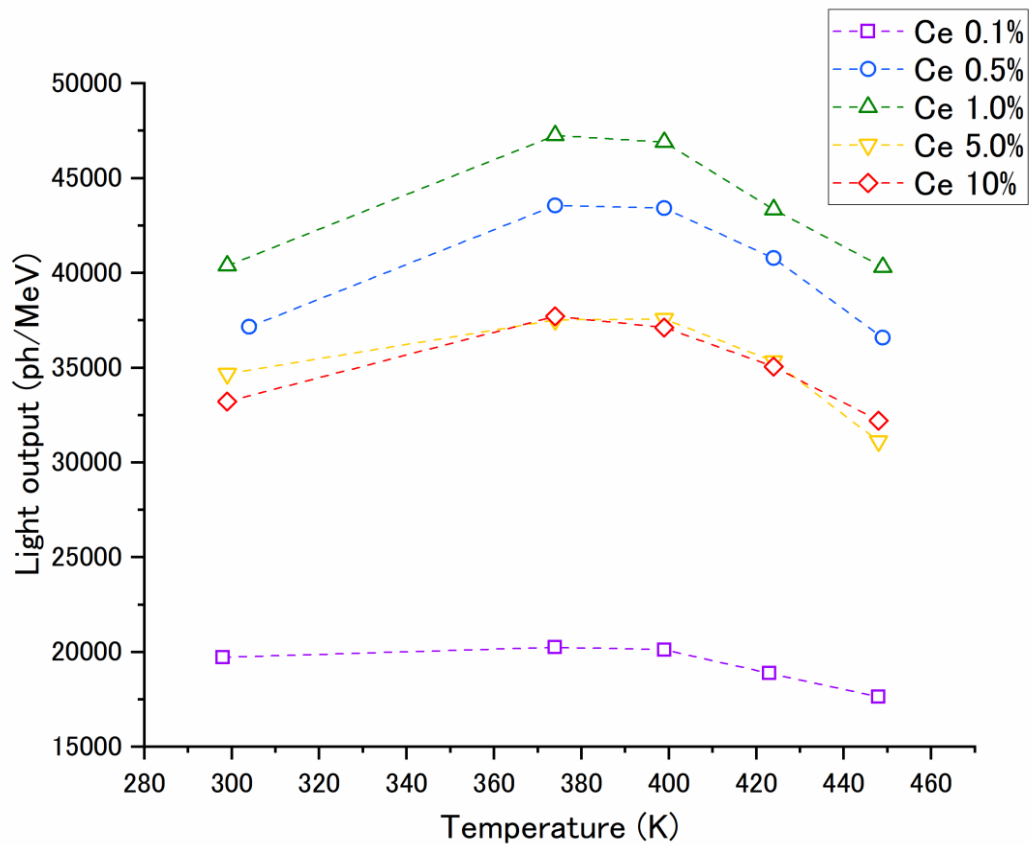

**Figure 11. Temperature dependence of gamma-ray photo-absorption peaks at various Ce concentrations, excited with $^{137}$Cs.**